## Journal of Geophysical Research: Space Physics

**Correspondence to:**
K. M. Laundal,
karl.laundal@ift.uib.no

# What is the appropriate coordinate system for magnetometer data when analyzing ionospheric currents?

**K. M. Laundal[1,2] and J. W. Gjerloev[1,3]**

[1]Birkeland Centre for Space Science, University of Bergen, Bergen, Norway, [2]Teknova AS, Kristiansand, Norway, [3]Johns Hopkins University Applied Physics Laboratory, Laurel, Maryland, USA

**Abstract** In this paper we investigate which coordinate representation is most appropriate when analyzing ground magnetometer data in terms of ionospheric currents, in particular the westward electrojet. The *AL* and the recently introduced *SML* index are frequently used as monitors of the westward electrojet. Both indices are based on ground magnetometers at auroral latitudes. From these magnetometers, the largest perturbation in the southward direction is selected as the *AL/SML* index at 1 min cadence. The southward component is defined as antiparallel to the orientation of the horizontal part of the Earths' main field, $\mathbf{B}_{0,H}$. The implicit assumption when using these indices as a monitor of the westward electrojet is that the electrojet flows perpendicular to $\mathbf{B}_{0,H}$. However, $\mathbf{B}_{0,H}$ is, in general, not perpendicular to the westward direction in coordinate systems that take nondipole terms of the Earth's magnetic field into account, such as apex and the Altitude Adjusted Corrected Geomagnetic coordinate systems. In this paper we derive a new *SML* index, based on apex coordinates. We find that the new index has less variation with longitude and universal time (UT), compared to the traditionally defined *SML*. We argue that when analyzing ionospheric currents using magnetometers, it is appropriate to convert the components to a corrected geomagnetic system. This is most important when considering longitudinal or UT variations, or when data from a limited region are used.

## 1. Introduction

Even before the discovery of the ionosphere, Kristian Birkeland realized that atmospheric currents at auroral latitudes could be described as a two-cell pattern in a coordinate system that is fixed with respect to the Sun [*Egeland and Burke*, 2010; *Birkeland*, 1901]. Since then, ground magnetometers have been routinely used to study currents flowing in the ionosphere and to monitor geomagnetic disturbance levels. Today, several hundred instruments measure magnetic disturbances throughout the world. In the SuperMAG collaboration [*Gjerloev*, 2009], data from more than 300 magnetometers are now available in a coherent and accessible format.

The institutions contributing to SuperMAG provide data in many different coordinate systems. Frequently encountered coordinate representations in studies of high-latitude geospace are (1) a geographic system (*XYZ*) where *X* is the horizontal magnetic field component toward geographic north, *Y* horizontal eastward, and *Z* downward; (2) a dipole coordinate system with the north component toward the centered dipole pole in the Northern Hemisphere; and (3) a compass-type coordinate system (*HDZ*), where *H* is the horizontal field strength, *D* is the declination, and *Z* is the vertical component. The dipole coordinate system is time varying in the reference frame of the Earth, since the position of the pole changes slowly as the main magnetic field of the Earth changes, while the other two are fixed.

SuperMAG converts the magnetometer data to a local magnetic coordinate system, which also varies with the Earth's main field. The northward component (denoted by *N*) in SuperMAG points along the typical hor-izontal direction of the magnetic field, determined using a 17 days sliding window (see details in *Gjerloev* [2012]). *Z* points down in this system, and *E* points in the *Z × N* direction. The benefit of this choice of coor-dinate system is that it is possible to derive from the data independently of the system chosen by the data providers, which for some magnetometers can be somewhat ambiguous.

None of the coordinate systems described above are perpendicular to contours of constant latitude and longitude in the Altitude Adjusted Corrected Geomagnetic (AACGM) and apex coordinates, which take nondipolar terms of the Earth's magnetic field into account. In this paper, we refer to such coordinate systems as corrected geomagnetic (CGM) coordinates. Since the Earth's magnetic field is not a perfect dipole, the contours of constant latitude and longitude in CGM coordinate systems are not necessarily perpendicular, and the coordinate system becomes nonorthogonal. Representing electrodynamic vector components in such systems, and subsequent calculations, is therefore not trivial. This is perhaps the reason why such systems are much more used for positional coordinates than for electrodynamic vector quantities [*Gasda and Richmond*, 1998]. However, since electrodynamic quantities are arguably better organized (more symmetrical) in CGM coordinates, conversion to CGM coordinates will likely make the quantities more invariant with respect to the longitude and hemisphere of the observation.

In this paper, we argue that when studying ionospheric currents, it is more appropriate to use a coordinate representation based on a corrected geomagnetic system, such as the apex systems [*Richmond*, 1995] and the AACGM system [*Baker and Wing*, 1989]. We particularly investigate how conversion of the north-south component affects the variation in UT and longitude of the *AL/SML* index. Since the 1960s, the *AL* index [*Davis and Sugiura*, 1966] has been used as an indicator of the westward auroral electrojet, which is the dominating source of ground magnetic perturbations in the auroral zone. The index is defined as the lower enveloping curve of the magnetic perturbations along the horizontal main field direction from around 12 magnetometers at auroral latitudes in the Northern Hemisphere. A similarly defined index, the *SML*, based on all magnetometers in SuperMAG between 40° and 80° magnetic latitude was recently introduced by *Newell and Gjerloev* [2011].

We compute an alternative *SML* index, based on the component pointing perpendicular to contours of constant latitude in the quasi-dipole (QD) coordinate system. This coordinate system, which is an apex system, was introduced by *Richmond* [1995], and we follow the procedure for vector component conversion prescribed in the same paper. The conversion involves both a rotation to an, in general, not orthogonal set of base vectors, and a scaling of the vector. The scaling accounts for variations in magnetic field strength, and varying density of contours of constant QD latitude seen in a geographic grid. The magnetic field component we use is always perpendicular to the east-west direction in QD coordinates, but it does not necessarily coincide with the north-south component, since the coordinate system is nonorthogonal.

The QD coordinates are defined in terms of magnetic field line tracing, along an International Geomagnetic Reference Field (IGRF) model line, from a point specified in geographic coordinates to the maximum height of the field line above the ellipsoidic Earth, $h_A$ (its apex). The QD latitude at the starting point is then defined by a mapping back to a spherical Earth ($R_E = 6371.009$ km) along a dipole field line:

$$\lambda_{qd} = \pm \cos^{-1} \left( \frac{R_E + h}{R_E + h_A} \right)^{1/2} \tag{1}$$

where $h$ is the height of the point where the tracing started. In this study we use $h = 0$ for all the magnetometers. The sign is determined by which hemisphere we map to. The longitude $\phi_{qd}$ is defined as the centered dipole longitude of the apex. This definition implies that corresponding coordinates in the two hemispheres belong to the same IGRF field line. QD circles of latitude and meridians from 2010 are shown for both polar regions in Figure 1. The QD grid is clearly not uniform seen in a geographic grid. The base vectors involved in converting the components of the magnetic field perturbations are described in more detail in the next section.

We use the code published by *Emmert et al.* [2010] to convert both the magnetometer position coordinates and the magnetic field components from geographic coordinates, taking secular variation of the main field into account.

Another apex coordinate system is the modified apex, which in place of $h$ in equation (1) has a reference height $h_R$. In contrast to QD coordinates, modified apex coordinates do not vary along IGRF magnetic field lines. Since we set $h = 0$ for all ground magnetometers, the coordinate system used in this paper is identical to modified apex coordinates with reference height 0. The definition of AACGM coordinates is very similar; instead of stopping the tracing at the apex, the tracing goes to the dipole equatorial plane. Because of the similarity, we expect that the arguments presented in this paper also hold for AACGM coordinates at polar latitudes.

In section 2 we present the technique used for converting the components, essentially repeating the description in *Richmond* [1995] and *Emmert et al.* [2010]. We also provide a geometric interpretation of the conversion, as well as a comparison with other coordinate systems when a westward current flows along a QD circle of latitude. In section 3 we present observational comparisons of the traditional *SML* index and a modified *SML* index based on QD *N*, the *QDSML*. Section 4 concludes the paper.

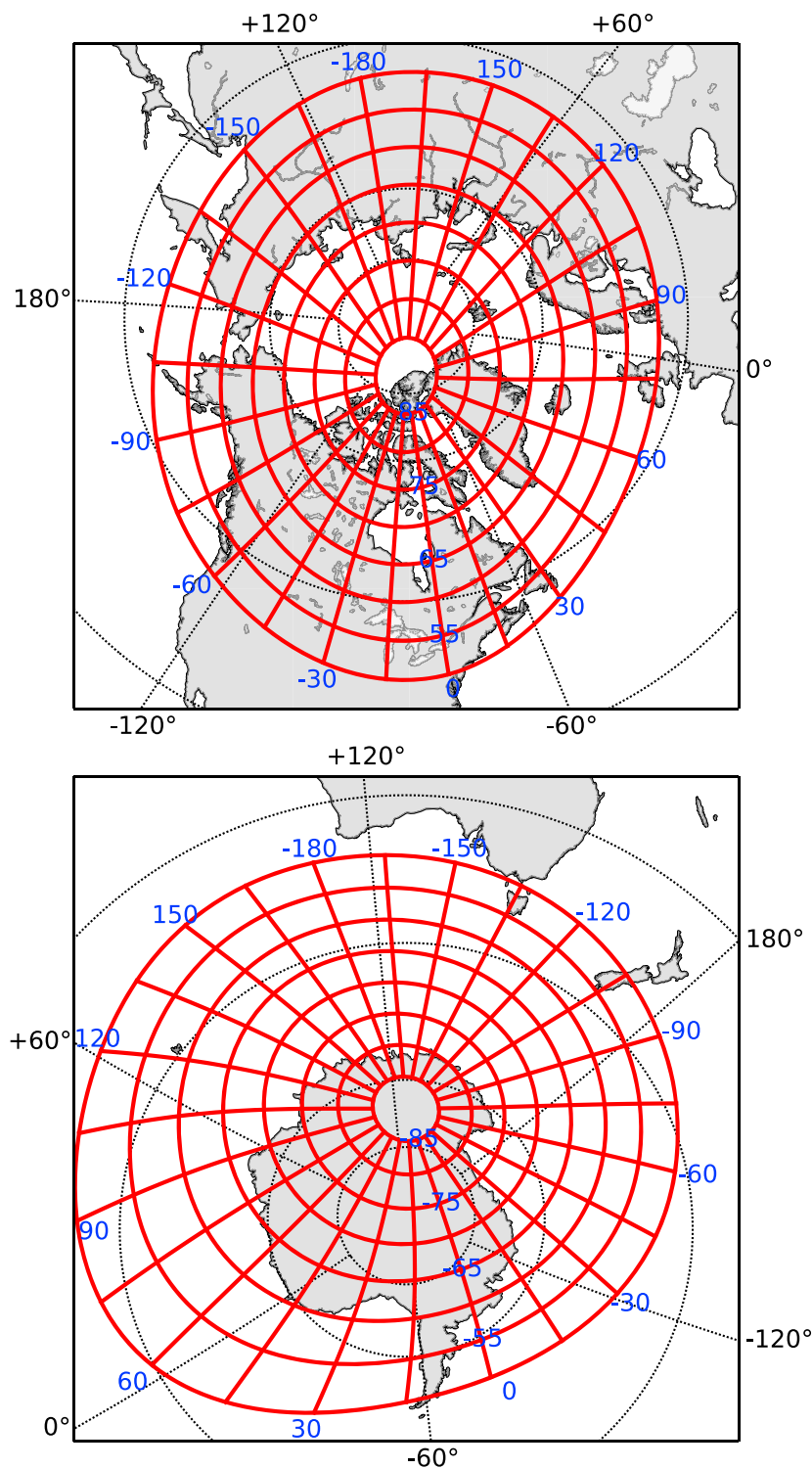

**Figure 1.** Quasi-dipole coordinates based on the IGRF 2010 coefficients plotted on a geographic grid in the polar regions. The circles of latitude extend to ±50°. The blue numbers correspond to QD latitude (along the 0° meridian) and longitude, and the black numbers correspond to geographic longitude. Both maps are centered at the QD poles and cover the same geographic area. Stereographic projection is used.

## 2. Technique

The data used for this study were obtained from the SuperMAG website. We use data from up to 137 magnetometers between 40 and 80 magnetic latitude (QD), spanning from 1990 to the end of 2009. The baselines have been subtracted by SuperMAG using a common method for all magnetometers [*Gjerloev*, 2009, 2012]. Note that SuperMAG originally uses AACGM coordinates for the magnetometer locations instead of QD. As explained above, the difference between these systems is small at high latitudes. In this paper we use only QD coordinates.

### 2.1. Magnetic Field Component Conversion

The conversion of the magnetic field components to QD coordinates involves two steps: First we need to convert from SuperMAG coordinates to geographic coordinates. To do this, we make the assumption that SuperMAG *N* aligns with the horizontal component of the magnetic field according to the (IGRF) model. With the IGRF expressed in geographic coordinates, it is then straightforward to rotate from the SuperMAG system to geographic coordinates. We have taken secular variations into account when making this rotation. To assess the validity of the assumption that the horizontal component of IGRF aligns with SuperMAG *N*, we have compared the declination computed by SuperMAG to the declination computed from IGRF for a selection of magnetometers. We have chosen to compare the average values from 2010 from the magnetometers that are part of both SuperMAG and Intermagnet. In total 106 magnetometers were compared. The average difference in declination was 0.4° and the median was 0.2°. Apart from one outlier, the largest difference was 2.4°. This suggests that the applied method works well for a large majority of observatories. The outlier was the Kiruna magnetometer where the difference in declination was 9.2°. It is likely that this difference comes from the local magnetic anomaly in Kiruna, which is not resolved by IGRF. Since anomalies are part of the measurements, they will, however, affect the orientation of the SuperMAG components. For the purpose of this study, we believe the benefit of a large number of magnetometers and a coherent preprocessing of the data outweighs a small number of errors in alignment.

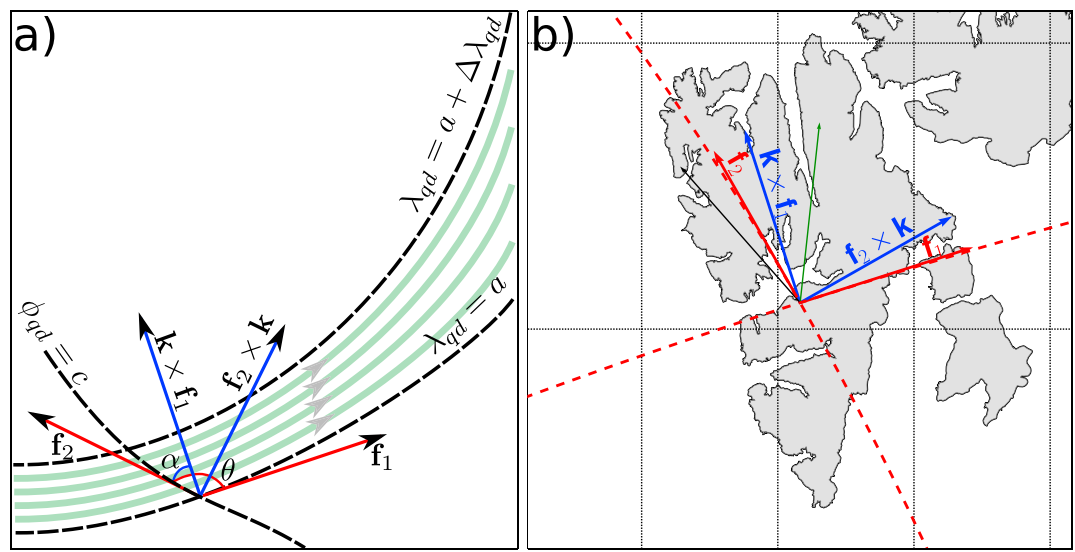

**Figure 2.** (a) Base vectors $\mathbf{f}_1$ and $\mathbf{f}_2$ and their horizontal normals. Two circles of latitude are shown, and one meridian (dashed curves). A zonal current is represented by green contours. The corresponding current density will be lower where the spacing between circles of latitude are larger; $\mathbf{k}$ is a unit vector pointing out of the plane. (b) Base vectors at the location of the Longyearbyen (LYR) magnetometer. Also shown are the direction of the IGRF horizontal field (green) and the north direction in dipole coordinates (black).

Next, we convert the components from geographic to QD coordinates. This is done using equations (7.12) and (7.13) in *Richmond* [1995]:

$$E_{QD} = \frac{\mathbf{f}_1 \cdot \Delta\mathbf{B}_{GEO}}{F} \tag{2}$$

$$N_{QD} = \frac{\mathbf{f}_2 \cdot \Delta\mathbf{B}_{GEO}}{F} \tag{3}$$

where $\mathbf{f}_1$ and $\mathbf{f}_2$ are base vectors defined by [*Richmond*, 1995, equations (6.6) and (6.7)]:

$$\mathbf{f}_1 = -(R_E + h)\mathbf{k} \times \nabla\lambda_{qd} \tag{4}$$

$$\mathbf{f}_2 = (R_E + h)\cos\lambda_{qd}\mathbf{k} \times \nabla\phi_{qd} \tag{5}$$

and $F = \mathbf{f}_1 \times \mathbf{f}_2 \cdot \mathbf{k}$ and $\mathbf{k}$ is a unit vector pointing upward; $h$ is the height at the magnetometer location, which we set to 0; $\mathbf{f}_1$ and $\mathbf{f}_2$ point tangential to contours of constant latitude and longitude, respectively. However, they are generally not of unit length or orthogonal, so the magnitude of the measured vector is not conserved in this conversion. The purpose of the conversion (and scaling) is to remove effects of local features in the Earth's internal magnetic field (nondipole terms in the IGRF). In the following, we give a geometric interpretation of the $N_{QD}$ component.

## 2.2. Geometric Interpretation

$N_{QD}$ is a projection of $\mathbf{B}$ on a unit vector in the $\mathbf{k} \times \mathbf{f}_1$ direction (perpendicular to currents that flow westward in QD coordinates), scaled by the geographic length per QD length along $\mathbf{k} \times \mathbf{f}_1$. To see this geometrically, we rewrite equation (3) (see also Figure 2):

$$N_{QD} = \frac{\mathbf{f}_2 \cdot \mathbf{B}_{GEO}}{F} = \frac{|f_2|}{|f_1||f_2|\sin\theta}\hat{f}_2 \cdot \mathbf{B}_{GEO}$$

$$= \frac{\hat{f}_2 \cdot \mathbf{B}_{GEO}}{|f_1|\sin\theta} = \frac{\hat{f}_2 \cdot \mathbf{B}_{GEO}}{\cos\alpha}\frac{1}{|f_1|} \tag{6}$$

where $\hat{f}_2$ is a unit vector in the $\mathbf{f}_2$ direction, $\theta$ is the angle between $\mathbf{f}_1$ and $\mathbf{f}_2$ and $\alpha$ is the angle between $\mathbf{f}_2$ and $\mathbf{k} \times \mathbf{f}_1$ (see Figures 2a and 2b). Because of the $\cos\alpha$ term, $\hat{f}_2 \cdot \mathbf{B}_{GEO} / \cos\alpha$ can be understood as the magnitude of $\mathbf{B}_{GEO}$ projected on a unit vector in the $\mathbf{k} \times \mathbf{f}_1$ direction. This component is then scaled by $1/|f_1|$ (this quantity is called the *width factor* by *Gasda and Richmond* [1998]). From equation (4) $|f_1|$ can be understood as a measure of QD length per geographic length along $\mathbf{k} \times \mathbf{f}_1$. Roughly speaking, when the density of QD circles of latitude (seen in a geographic frame) is high, $|f_1|$ will be large, when the density is low, $|f_1|$ will be small. The scaling thus compensates for variations in latitudinal extent over which zonal (in QD) ionospheric currents are spread. This variation in latitudinal extent is a consequence of the Earth's nondipolar magnetic field. $\nabla\lambda_{qd}$ and $\nabla\phi_{qd}$ are generally large where the magnetic field is strong, and low where the magnetic field is weak.

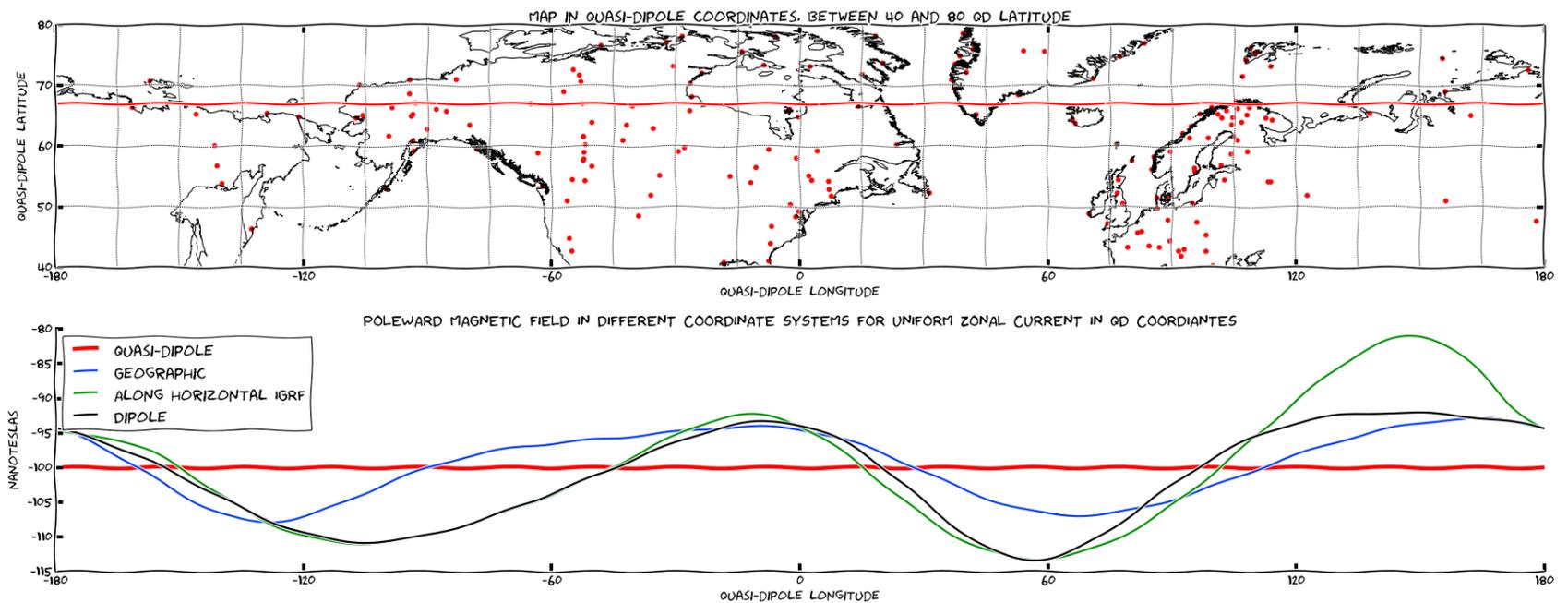

**Figure 3.** (top) Equirectangular projection of the 40–80 QD latitude region. The coastlines are warped to QD coordinates. The red dots show the locations of the magnetometers in the SuperMAG collaboration used in this study. The red line shows the QD latitude (67°) of a constant zonal westward current. (bottom) The poleward components, seen in four different coordinate systems, of the magnetic field from a uniform westward current in QD coordinates.

Figure 2a shows a sketch of $\mathbf{f}_1$ and $\mathbf{f}_2$ at $(\lambda_{qd}, \phi_{qd}) = (a, c)$. The green contours represent a uniform iono-spheric current in the QD eastward direction, along $\mathbf{f}_1$. The horizontal ground magnetic perturbation from this current will be along $\mathbf{k} \times \mathbf{f}_1$. In QD coordinates, the current density is constant, but from a geo-graphic grid, the current density varies with the spacing of contours of constant $\lambda_{qd}$. The vector component conversion compensates for this nonuniformity of the QD coordinates.

The same argument as above holds for $E_{QD}$ and meridional currents. The conversions in equations (3) and (2) thus converts the magnetic field components so that they can more directly be interpreted in terms of currents flowing along contours of constant QD latitude or longitude.

Figure 2b shows base vectors for the Longyearbyen (LYR) magnetometer. The direction of $\mathbf{k} \times \mathbf{f}_1$ (which is perpendicular to the contour of constant QD latitude) is significantly different from the direction of the IGRF horizontal field (shown in green), which presumably coincides with SuperMAG $N$, as well as the north vector in a centered dipole coordinate system (shown as a black arrow).

Consulting Figure 1, it can be seen that in the Northern Hemisphere $\nabla\lambda_{qd}$ is larger at the Atlantic and Pacific longitudes compared to Asian and North American longitudes and that $\nabla\phi_{qd}$ must follow the opposite pat-tern. Mercator and stereographic projections are used in Figures 2b and 1, respectively. Both projections preserve local angles.

### 2.3. Synthetic Example

Above we argue that the traditional magnetic field component representations are not invariant with respect to longitude and hemisphere if currents are organized in corrected geomagnetic coordinates. Let us assume that the electrojet flows strictly westward in QD coordinates. Figure 3 shows the northward compo-nent in four different coordinate systems, from such a current, at 67° QD latitude, producing a perturbation in the QD coordinate system in the northward direction of −100 nT (by definition). This magnetic field per-turbation is then converted to geographic (blue), using the inverse operation of equations (2) and (3) [i.e., *Richmond*, 1995, equation (8.8)]. The geographic vector is then further converted to dipole coordinates (black) and along the IGRF horizontal field (green). We see that $N_{QD}$ is stronger compared to the north-ward component in the other coordinate systems in the eastern part of North America, and in Russia. It is weaker than the other components at European/Atlantic longitudes and at Pacific longitudes. In the latter regions, the spacing between circles of latitude is smaller so that the current density becomes stronger. The measured vectors in these regions will therefore be scaled down when converting to QD coordinates.

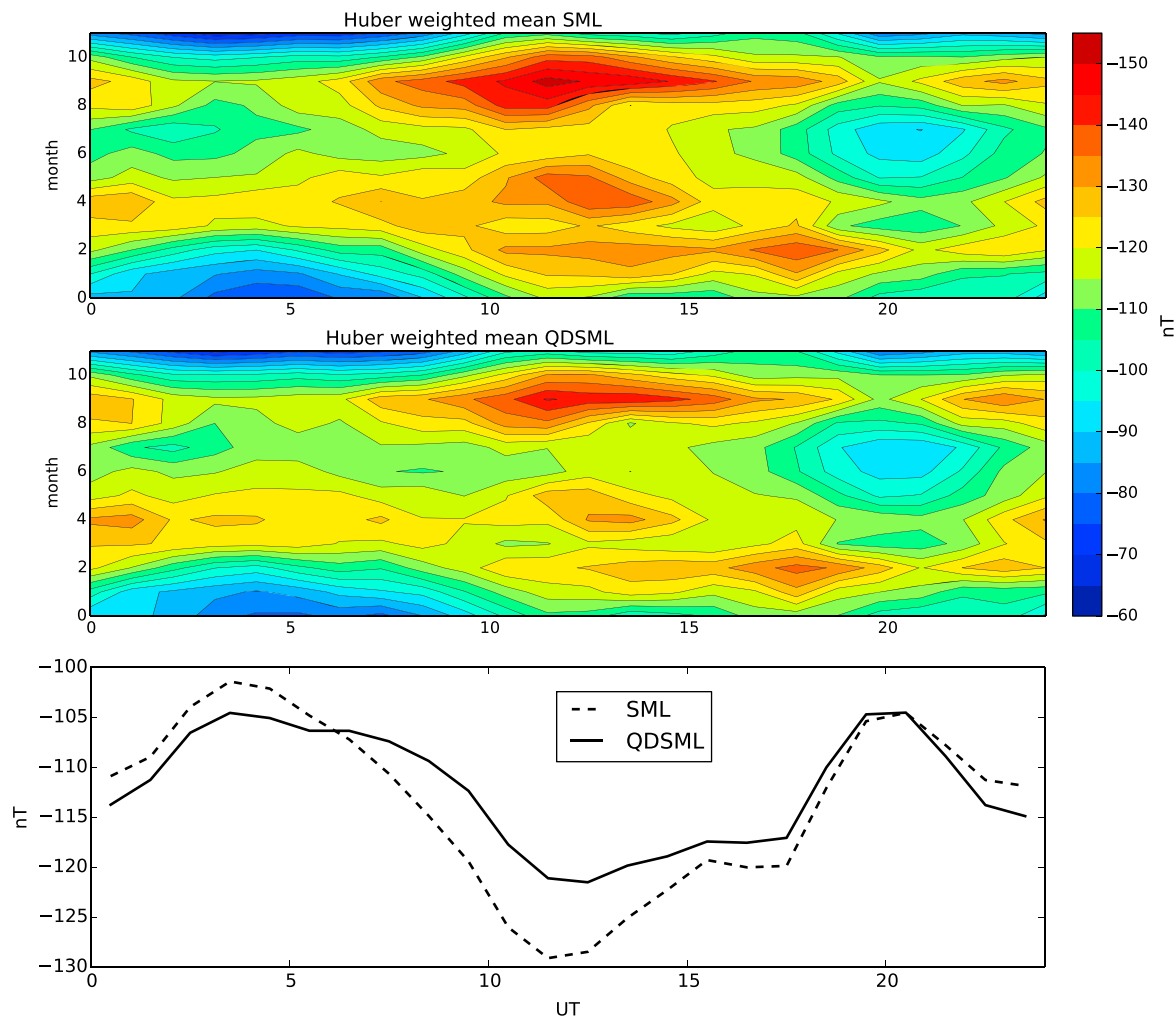

**Figure 4.** The mean (top) *SML* and (middle) *QDSML* binned by UT (*x* axis) and month (*y* axis). (bottom) Curves that are summed over the months in the contour plots.

The map is also shown in QD coordinates, with the coastlines warped to this system. The red dots show magnetometers in the SuperMAG collaboration. Even though the number of magnetometers is large, they clearly cover the North American and European sectors better than the Russian sector.

This example illustrates that if we represent the magnetic field components in terms of a magnetic dipole or geographic grid, or along $B_{H,IGRF}$, we will observe longitudinal variations that are purely a result of local features in the main field. When studying effects of externally induced disturbances, such a variation is generally not what we are interested in. The above example is of course synthetic, and it is relevant only if ionospheric currents really are better organized in QD/apex coordinates. In the following we compute the SuperMAG $N$ ($N_{SM}$) and $N_{QD}$ components for a large data set, in order to demonstrate the effect with observations.

## 3. Observations

We have used the conversion method described above to compute $N_{QD}$ for the years 1990 to the end of 2009. In the conversion, secular variation is taken into account so that $f_1$ and $f_2$ change slowly as the Earth's main field changes. We computed a modified *SML* index, the *QDSML*, using the minimum value of $N_{QD}$ between all stations between QD latitude 40 and 80. We also computed the *SML* index using the $N_{SM}$ component, as reference. The magnetometer coverage through these years is variable, with a minimum number of 32 (only in 1990), and a maximum of 137. The median number of magnetometers was 92. In total, > 923 million magnetometer data measurements (at 1 min cadence) were used to derive the indices, producing > 10 million *SML* and *QDSML* values.

To reduce the sensitivity to errors (e.g., spikes), we use an iterative scheme with Huber weighting [*Huber*, 1964] when computing the mean and standard deviation. Specifically, we start by computing the mean and standard deviation of the raw data. Then we introduce a weight defined by

$$w_i = \min \left\{ 1, \frac{k}{|x_i - <x_i>|/\sigma} \right\} \qquad (7)$$

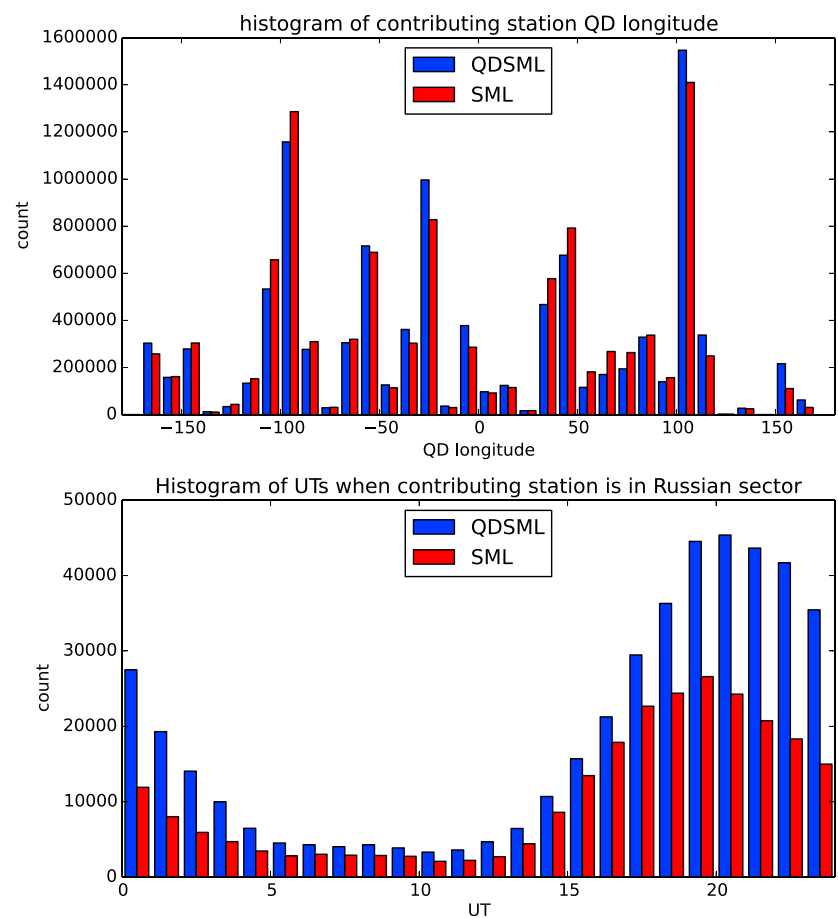

**Figure 5.** (top) Histogram of the quasi-dipole longitude of the station contributing to *SML* (red) and *QDSML* (blue). (bottom) Histogram of the UT hour when the contributing station was at QD longitude > 140°, which is sparsely covered by magnetometers.

where $\sigma$ is standard deviation and the Huber coefficient $k$ was set to 1.5, reducing the weights of points that are more than 1.5 standard deviations from the mean. Then the mean and standard deviation was computed again using these weights. New weights are then computed based on the new mean and standard deviation. The process is repeated until the difference between consecutive means are less than 0.1 nT. Comparisons with other techniques show that the Huber means generally are less than the simple mean, but larger than the median, in absolute value. Except for the scale, the statistics presented below give similar results with all three techniques.

For the whole 20 years data set, the mean *SML* was −112.7 nT, and the standard deviation was 79.3 nT. For the *QDSML*, the mean value was −111.3 nT and standard deviation was 78.4 nT. These values are very similar, which suggests that the longitudinal variation to some extent is averaged out when using the full global data set from all UTs.

In the following we present a comparison of (1) the UT dependence of the *SML* and detected substorm onsets and (2) superposed epoch curves of substorms observed at different longitudes.

### 3.1. *SML* and Substorm Onset UT Dependence

*Singh et al.* [2013] identified a clear UT dependence in the *SML* index, and in the frequency of substorm onsets detected using the *SML* index and an algorithm developed by *Newell and Gjerloev* [2011]. Their results were consistent with an earlier study, based on the *AL* index [*Lyatsky et al.*, 2001]. A UT dependence in indices such as *AL* and *SML* is expected from the nonuniform distribution of magnetometers. Another expected UT dependence is due to the changing dipole tilt angle with the solar wind flow [*Cliver et al.*, 1990]. In addition, *Lyatsky et al.* [2001] argue that there is an effect of ionospheric illumination on the probability of substorm onsets. However, previous studies have not used magnetic field components converted to CGM coordinates. Since the maximum of the westward electrojet on average is located in the postmidnight region, the longitudinal variation described in section 2 will translate to a UT dependence. We thus expect that part of the UT dependence observed by *Singh et al.* [2013] and *Lyatsky et al.* [2001] is due this longitude dependence. In this section we do the same analysis as *Singh et al.* [2013], using both the *SML* and the *QDSML* index.

In Figure 4 we compare the diurnal and seasonal variations observed in the *SML* and *QDSML* indices. The data were binned according to month and UT hour, and means were calculated in each bin. The result is plotted as contour plots with common color scale. The *SML* shows a very clear maximum at equinox and at UTs between 8 and 15, in accordance with *Singh et al.* [2013]. The quasi-dipole *SML* also shows maxima, but they are much less pronounced. The lower plot shows only the UT dependence, computed by integrating over seasons. This plot clearly shows that the UT variation is reduced with the *QDSML* index. The most prominent discrepancy from a uniform distribution is the minimum amplitude around 20 UT (appears as a maximum, since westward currents produce negative perturbations). This minimum is very likely due to the sparse coverage of magnetometers in the Russian sector.

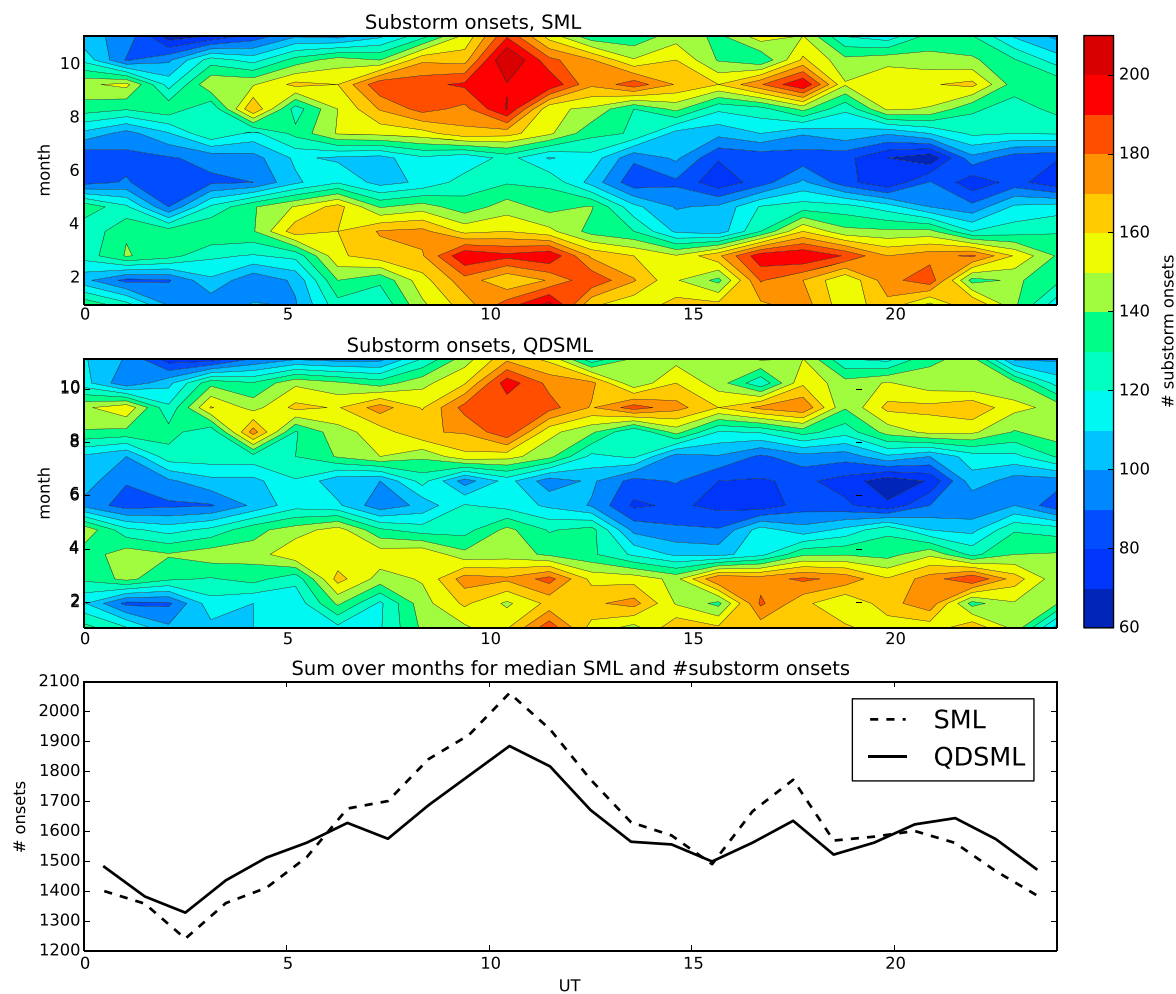

**Figure 6.** Substorm onset count binned by UT ($x$ axis) and month ($y$ axis) for (top) *SML* and (middle) *QDSML*. (bottom) Curves that are summed over the months in the contour plots.

To analyze the effect of magnetometer distribution further, we show a histogram of which longitudes contribute to *SML* and *QDSML* in Figure 5 (top). The red (blue) bars show the number of data points that magnetometers from a given 10° wide longitude sector contributes to the *SML* (*QDSML*) index. This figure shows that the distribution of magnetometers is clearly nonuniform. We also see that the distribution of the data points is different between *SML* and *QDSML* and that the change roughly follows the ratio between $N_{SM}$ and $N_{QD}$ in Figure 3.

The lower part of Figure 5 shows a histogram of the UT when the stations at > 140° QD longitude (western Russia) contributed to *SML* (red) and *QDSML* (blue). We see a clear maximum around 20 UT. From this we conclude that the western electrojet has its maximum amplitude at > 140° QD longitude at this time. The minimum *SML* and *QDSML* amplitudes at 20 UT in Figure 4 can therefore probably be attributed to the sparse magnetometer coverage in this region. Interestingly, the minimum is not very different between *SML* and *QDSML*, despite the fact that *QDSML* contains about twice as many data points from the > 140° QD longitude sector as *SML*. We speculate that this has to do with the latitudinal distribution of the magnetometers here. If the magnetometers are located outside the average maximum latitude of the westward electrojet, they will on average measure lower amplitudes.

Figure 6 shows the frequency of substorm onsets using *SML* (top) and *QDSML* (middle). The algorithm of *Newell and Gjerloev* [2011] dictates that a substorm onset is identified at $t = t_0$ (in minutes) when

$$SML(t_0 + i) - SML(t_0) < -15i \text{ nT}, \qquad i = 1, 2, 3 \tag{8}$$

$$\sum_{i=4}^{i=29} SML(t_0 + i)/26 - SML(t_0) < -100 \text{ nT} \tag{9}$$

with the additional requirement that no substorm onsets must have occurred the previous 20 min.

With *SML*, 38,512 substorms were detected, and 37,966 substorms with *QDSML*. Figure 6 shows that the UT dependence is clearly reduced when the converted component is used. In Figure 6 (bottom) we see that the peaks in substorm count detected by $SML$ at $\approx$ 10 and $\approx$ 17 UT are significantly reduced in $QDSML$. It is likely that the difference between $QDSML$ and $SML$ at these times is due to the downscaling of $N_{QD}$ compared to $N_{SM}$ in regions that are densely covered by magnetometers, such as the western part of North America and in Greenland (Figure 3). The combination of downscaling $N_{SM}$ in densely covered regions and upscaling $N_{SM}$ (when converting to QD) in sparsely covered regions such as Russia probably also explains the overall reduction in substorm count with $QDSML$ compared with $SML$.

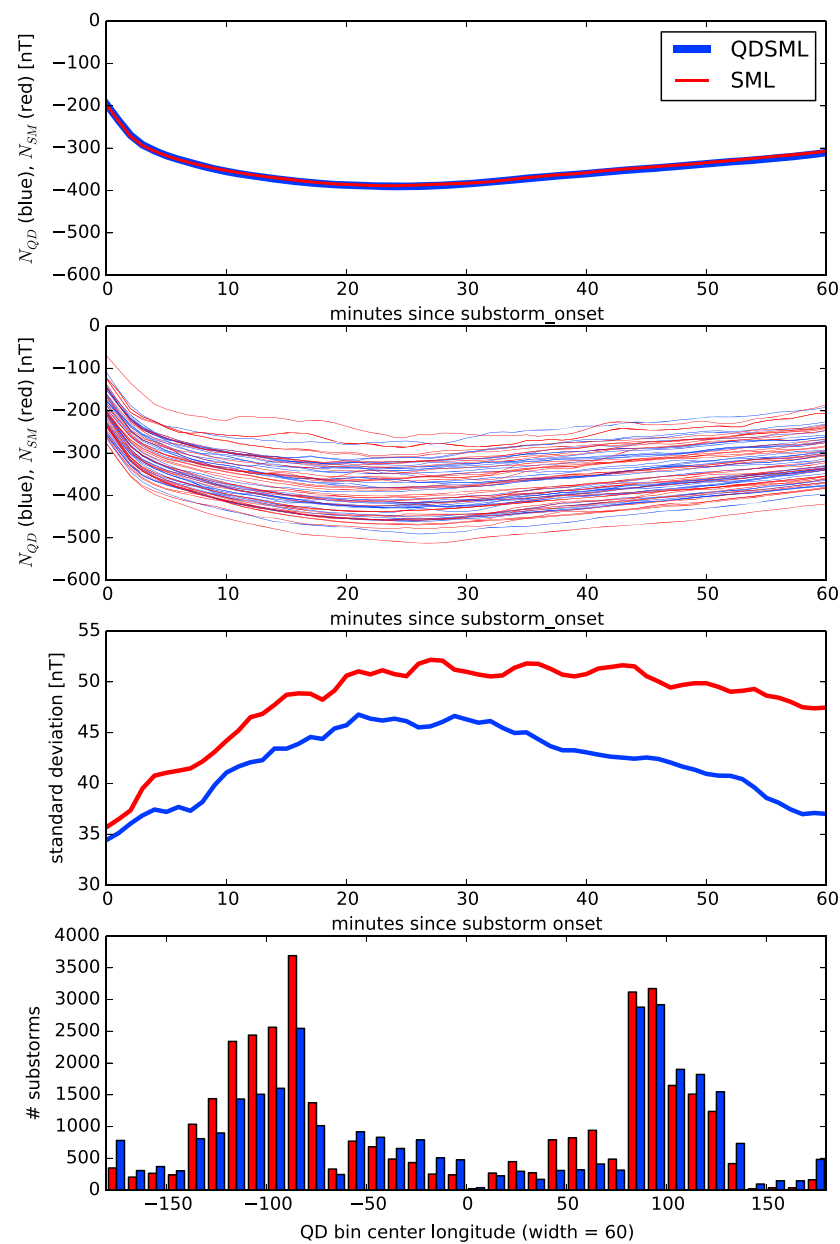

### 3.2. Longitudinal Variations in Magnetic Signatures of Substorm Dynamics

The longitudinal variation in $N_{QD}$ compared to $N_{SM}$ also means that localized phenomena will have a longitudinal variation when corrected geomagnetic components are not used. One such phenomenon is substorms. We find that among the 38,512 (37,966) substorms detected in $SML$ ($QDSML$), 84% (81%) had contributions only from stations separated by less than 60° QD longitude (4 h magnetic local time) during the first hour. We now look at how the average substorm development differs between longitude sectors when represented in QD and SM coordinates.

**Figure 7.** (first row) Superposed epoch of substorms, using $N_{SM}$ (red) and $N_{QD}$ (blue). The lines are largely overlapping. (second row) Superposed epoch of substorms binned by longitude (the first hour detected entirely within 60° QD longitude bins). Thirty-six bins are shown for $QDSML$ (blue) and $SML$ (red). (third row) The standard deviation of the above curves as a function of time since substorm onset for $QDSML$ (blue) and $SML$ (red). (fourth row) Histogram with substorm count in each bin.

Superposed epoch curves for all substorms (not sorted by longitude) are shown in Figure 7 (first row). We see that these curves are almost identical, supporting the idea that longitudinal variations between the coordinate representations are averaged out when using large global data sets. The thin lines in Figure 7 (second row) show superposed epoch curves for substorms which have contributions only from magnetometers in 60° wide longitude bins. The bins are computed in steps of 10° (36 bins). We see that there are significant variations between different longitude bins for each coordinate representation. Much of this variation is probably due to differences in latitudinal distribution of magnetometers at different longitudes.

To quantify the variation with longitude and to compare the northward component representations, we have computed the standard deviation among the superposed epoch curves. This is shown in Figure 7 (third row). Here we clearly see that the substorms where $N_{QD}$ is used have less variation than with $N_{SM}$. The standard deviation is reduced by on average 12%. Without Huber weights the standard deviation is reduced by on average 25%. However, weighting removes much of the contribution from data points with very few substorms, which could be dominated by extreme events. Figure 7 (fourth row) shows the number of substorms in each QD longitude bin as a bar plot. We see that the distribution of detected onsets become more uniform with longitude with QD coordinates. However, the number of detected onsets is still much higher at North American and European longitudes, where the density of magnetometers is high.

These results imply that when single magnetometer arrays from a confined region, such as International Monitor for Auroral Geomagnetic Effects (IMAGE), which mainly covers the Scandinavian region, or Canadian Array for Realtime Investigations of Magnetic Activity (CARISMA) in North America, we must expect different results because of differences in the Earth's main magnetic field. If corrected geomagnetic components are used, these differences will be reduced. Remaining differences can more directly be interpreted in terms of other effects such as magnetometer distribution, differences in ionospheric conductance, local differences in induced currents in the ground, and neutral dynamics.

## 4. Discussion and Conclusions

In a ground state (without the solar wind), the magnetic field around Earth would be almost entirely determined by the currents flowing in its core, and it would be similar to the IGRF at large distances. At large distances, this field is dominated by the dipole term. To a large extent, a dipole is what the solar wind interacts with. The disturbances created by this interaction become increasingly affected by higher-order terms in the spherical harmonic representation of the Earth's magnetic field as they are mapped to lower altitudes. This creates distortions and asymmetries which we remove by converting to corrected geomagnetic coordinates. Details on how to express electric fields, magnetic fields, currents, and velocities in apex coordinates are given by *Richmond* [1995] and discussed by *Gasda and Richmond* [1998].

In this paper we have shown that when magnetic perturbation vectors are converted to quasi-dipole (apex) coordinates, according to the method prescribed by *Richmond* [1995] and *Emmert et al.* [2010], the longitudinal variation in the *SML* index is reduced.

### 4.1. Decrease in UT Variation

The decrease in UT variation observed when repeating part of the study done by *Singh et al.* [2013] is purely due to an essentially geometric correction of the data. Our results diminish the evidence for an ionospheric influence on substorm triggering. However, there is still a variation which leaves questions open for further research. We speculate that the main contribution to the remaining variation comes from the nonuniform distribution of magnetometers [*Ahn et al.*, 2000], and from the fact that at a fixed magnetic local time/magnetic latitude, the ionospheric conductance will vary with UT, since there is an offset between the magnetic and geographic poles.

The implicit assumption when converting the components to QD coordinates, which is partly qualified by the data analysis in this paper, is that ionospheric currents are better organized in a corrected geomagnetic system. It is, however, important to be aware that the ionospheric currents are not purely an effect of external forcing and that thermospheric feedback (and, mainly at lower latitudes, dynamo effects) also plays a role [e.g., *Förster et al.*, 2008]. Thermospheric dynamics is to first-order organized in geographic coordinates and geographic local time. Interaction between the thermosphere and the charged ionosphere is therefore expected to show up as a UT or longitudinal variation when QD components are used. However, interpretation of such variations becomes more straightforward when this conversion is done, since the signal is not mixed with variations in the main field. Thermosphere/ionosphere coupling may be another explanation for the remaining variation that we observe in the data.

### 4.2. Invariance With Respect to Longitude and Hemisphere

As indicated by Figure 1, the Southern Hemisphere has an even stronger longitudinal variation at certain latitudes. A similar westward current as in Figure 3, only at $-67°$ QD latitude, would give a QD $N$ component that is up to $\sim 60\%$ stronger than $N_{GEO}$, and more than 40% stronger than $N_{SM}$. In fact, even closer to the pole (e.g., at 80° QD latitude), the orientation of $N_{QD}$ points more or less opposite to both $N_{SM}$ and $N_{GEO}$ at certain longitudes. The large differences between longitude and hemispheres also indicates that results of interhemispheric comparisons of magnetometer measurements will be highly dependent on the choice of coordinate representation.

Clearly, the interpretation of magnetometer measurements is highly dependent on the choice of coordinate system. One example is shown in Figure 8, from the Longyearbyen magnetometer (see base vectors at this location in Figure 2). Here we see the $N$ component in SM (red), QD (blue), and GEO (black) coordinates. While the overall development is similar, there are clear differences in detail between the QD component and the others. The maximum difference in magnitude between $N_{QD}$ and $N_{SM}$ is 906 nT. We note that the ratio between the different components is not constant, since the orientation of the disturbance magnetic field vector varies.

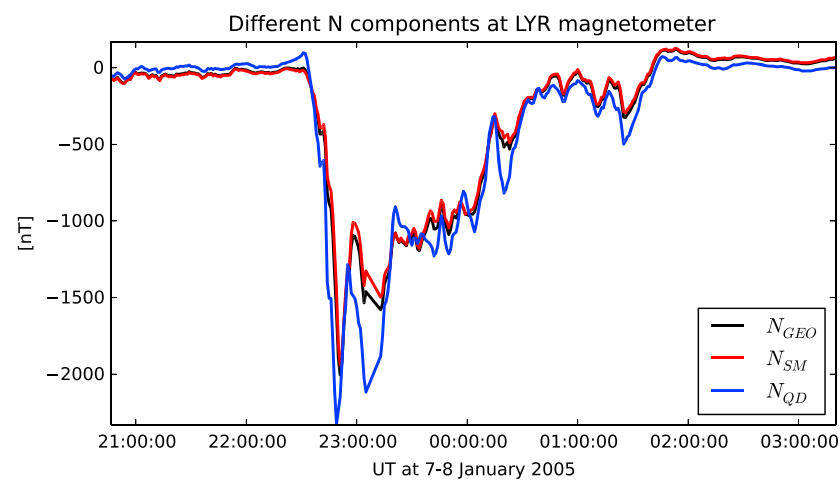

**Figure 8.** Example plot of the magnetic perturbation $N$ component at the Longyearbyen (LYR) magnetometer for three different coordinate systems.

### 4.3. Conclusions

Quasi-dipole magnetic field components represent the magnetic field in directions that are perpendicular to QD contours of constant latitude and longitude and compensates for variations in the Earth's main field.

We have demonstrated that ground perturbation magnetic field measurements represented in quasi-dipole coordinates (1) has less variation with UT and longitude and (2) are more invariant to longitude and hemisphere compared to traditional coordinate representations. We therefore recommend using corrected geomagnetic coordinate components when analyzing currents in space. The conclusions in this paper also applies to measurements from low-Earth orbit.

**Acknowledgments**
This study was supported by the Research Council of Norway/CoE under contract 223252/F50. The data for this study were downloaded from the SuperMAG website: http://supermag.jhuapl.edu/. For the ground magnetometer data we gratefully acknowledge the following: Intermagnet; USGS, Jeffrey J. Love; CARISMA, PI Ian Mann; CANMOS; The S-RAMP database, PI K. Yumoto and K. Shiokawa; The SPIDR database; AARI, PI Oleg Troshichev; The MACCS program, PI M. Engebretson, Geomagnetism Unit of the Geological Survey of Canada; GIMA; MEASURE, UCLA IGPP and Florida Institute of Technology; SAMBA, PI Eftyhia Zesta; 210 Chain, PI K. Yumoto; SAMNET, PI Farideh Honary; the institutes who maintain the IMAGE magnetometer array, PI Eija Tanskanen; PENGUIN; AUTUMN, PI Martin Conners; DTU Space, PI Jürgen Matzka; South Pole and McMurdo Magnetometer, PIs Louis J. Lanzarotti and Alan T. Weatherwax; ICESTAR; RAPIDMAG; PENGUIn; British Artarctic Survey; McMac, PI Peter Chi; BGS, PI Susan Macmillan; Pushkov Institute of Terrestrial Magnetism, Ionosphere, and Radio Wave Propagation (IZMIRAN); GFZ, PI Monika Korte.

Alan Rodger thanks Susan Macmillian and another reviewer for their assistance in evaluating the paper.